\documentstyle[aps,prl,multicol,amsfonts,amssymb,epsfig]{revtex}

\newcommand{\Eq}[1]{Eq.~(\ref{#1})}

\newcommand{\vev}[1]{\langle #1 \rangle}

\renewcommand{\Im}{{\rm Im}}

\newcommand{\prt}{\partial}

\renewcommand{\v}[1]{{\bbox{ #1}}}
\renewcommand{\t}[1]{{\tilde #1}}
\newcommand{\up}{\uparrow}
\newcommand{\down}{\downarrow}
\renewcommand{\dag}{\dagger}

\newcommand{\al}{\alpha}
\newcommand{\bt}{\beta}
\newcommand{\del}{\delta}
\newcommand{\Del}{\Delta}
\newcommand{\eps}{\varepsilon}

\newcommand{\Ga}{\Gamma}

\newcommand{\om}{\omega}

\begin{document}
\draft
\widetext

\title{ Theory of Quasi-Particles in the Underdoped High $T_c$ Superconducting
State }

\author{ Xiao-Gang Wen and Patrick A. Lee  }
\address{Department of Physics,
Massachusetts Institute of Technology,  Cambridge, Massachusetts 02139}

\maketitle

\widetext
\begin{abstract}
\rightskip 54.8pt

The microscopic theory of superconducting (SC) state in the  $SU(2)$ slave-boson
model is developed. We show how the pseudogap and Fermi surface (FS) segments
in the normal state develop into a $d$-wave gap in the superconducting state.
Even though the superfluid density  is of order $x$ (the doping concentration),
the physical properties of the low lying quasiparticles  are found to resemble
those
in BCS theory. Thus the microscopic theory lay the foundation for our
earlier phenomenological discussion of
the unusual SC properties in the underdoped cuprates.

\end{abstract}


\begin{multicols}{2}

\narrowtext

It has become clear in the past several years that the
underdoped cuprates show many highly unusual properties compared to
conventional metals/superconductors,
both in the normal and SC states.
The most striking of all are the pseudo spin-gap in the normal
state and the low superfluid density (of order $x$).
The photoemission experiments\cite{4} reveal that the pseudogap is of the same
size and
$\mbox{\boldmath $k$}$ dependence as the $d$-wave SC gap.  Furthermore,
the pseudogap is essentially independent of doping and the SC
transition temperature $T_c$ (which is proportional to $x$)
can be much less then the pseudogap in the low doping limit.
A phenomenological model was developed to described the above
unusual SC properties.\cite{LeeWen} The model is based on two
basic assumptions: A) the  superfluid density is given by $x$, and B)
the quasiparticle (qp) dispersion in presence of an external electromagnetic
gauge potential has a BCS form:
\begin{equation}
 E^{(sc)}_{\v A}(\v k)=\pm \sqrt{\eps^2(\v k)+
 \Delta^2(\v k) } - \frac{\v A}{c} \cdot {\v j}({\v k})
 \label{Esc}
\end{equation}
where ${\v j}(\v k)$ is the current carried by the ``normal state qp'' with
momentum ${\v k}$.  In ref. \cite{LeeWen} ${\v j}$ is assumed to be
$-e{\v v}_F = -e \partial_{\v k} \eps$.  With these assumptions the model
successfully explains the observations that linear temperature dependence of
the superfluid density is independent of $x$ and that $T_c \approx x\Delta_0$,
a strong violation of the BCS ratio.

It was recently pointed out\cite{Millis} that in conventional BCS
superconductors
developed out of a Fermi liquid, the Fermi liquid correction to the qp current
appears, so that in general\cite{Larkin,Leggett}
\begin{equation}
j({\v k}) = -e \alpha {\v v}_F \;\;\; .
\label{Eq.2}
\end{equation}
For example, if only a single Fourier component of the Landau parameter
$F_{1s}$ is important, $\alpha = 1 + F_{1s}/3$, but more complicated anisotropic
Landau parameters are generally possible.  With the more general assumption
\Eq{Eq.2}, the phenomenological model now predicts that
\begin{equation}
\frac{\rho_s(T)}{m} = \frac{x}{ma^2} - \frac{2\ln2}{\pi} \alpha^2
\left( \frac{v_F}{v_2} \right) T
\label{Eq.3}
\end{equation}
where $v_2$ is the velocity of the $d$-wave SC qp in the direction perpendicular
to ${\v v}_F$.  We have seen that in order to agree with experiments, $\alpha$
near the nodes is either exactly unity or close to it, and must be independent
of $x$.  On the other hand, if one attempts to describe the normal state of
underdoped cuprates by Fermi liquid theory, one faces the dilemma that the area
of the Fermi surface is $1-x$ while the spectral weight of the Drude peak
(which develops into the superfluid density in the SC state) is proportional to
$x$.  In Fermi liquid theory this can be accommodated by assuming $1 +
F_{1s}/3 =
x$.  From \Eq{Eq.3} we see that within this scenario, the $T$ dependence of
$\rho_s$ is too small by a factor of $\alpha^2 = x^2$.  Thus a proper
microscopic theory must explain in a natural way why the spectral weight is $x$
while $\alpha \approx 1$.  We believe this requirement is a central issue in the
high $T_c$ problem, and lies at the heart of the debate of spin-charge
separation\cite{Ands} vs Fermi liquid theory in the normal state.

In this paper we show that this requirement is satisfied by the $SU(2)$
slave-boson theory.\cite{WL,LNNW} The slave-boson theory was developed to
satisfy the constraint of no double occupation in the $t$-$J$ model.  The
electron is decomposed into a fermion and a boson and naturally incorporates the
physics of spin-charge separation in the normal state. The charge is
carried by $x$ bosons so that assumption A is automatic. The difficulty is that
at the mean field (MF) level, the  SC state is described by the condensation of
slave bosons and the SC qp dispersion is given by the fermion dispersion. Since
$\v A$ couples directly only to the bosons, the shift in the qp spectrum is
reduced and  in \Eq{Eq.2}, $\alpha$ is less than one.  In fact, in the
traditional $U(1)$ formulation, $\alpha = x$ and this theory faces the same
difficulty as Fermi liquid theory.

We recently introduced a new formulation of the slave-boson theory, the $SU(2)$
theory\cite{WL,LNNW}, which incorporates an $SU(2)$ symmetry that is known
to be important at half-filling.  In that case a fermion doublet $\psi^T =
(\psi_\uparrow,\psi_\downarrow^\dag)$ was introduced because in the projected
subspace, both $\psi_\uparrow$ and $\psi_\downarrow^\dag$ represent the removal
of an up-spin.  We extended this symmetry to finite $x$ by introducing a boson
$SU(2)$ doublet $b^T = (b_1,b_2)$.  The spin-up and spin-down electron
operators are given by the $SU(2)$ singlets
$c_\uparrow(i) = \frac{1}{\sqrt{2}}b^\dag(i)\psi(i) \;\; , \;\;
c_\downarrow(i) = \frac{1}{\sqrt{2}}b^\dag(i)\bar{\psi}(i) $
where $\bar{\psi}=i\tau^2\psi^\ast$.  The advantage of this formulation is that
near half-filling, low lying fluctuations which were ignored in the $U(1)$
formulation are included at the MF level.  Furthermore, we found that if we go
beyond MF theory and include an attraction between the fermions and bosons, the
electron spectral function contains qp-like peaks which exhibit a gap near
$(0,\pi)$ but appear gapless in a finite region near
$\left(\frac{\pi}{2},\frac{\pi}{2}\right)$, leaving what we may term a FS
segment.  These features are in qualitative agreement with the photoemission
experiments.  It is then natural for us to extend the same treatment to the SC
state, and see how the FS segment evolves into $d$-wave SC qp.  Then we can
study the coupling of these qp to ${\v A}$ in order to justify assumption B
and determine whether $\alpha = 1$

The $SU(2)$ slave-boson model is described by the following effective theory
at MF level (for details see Ref. \cite{WL,LNNW}):
$H_{mean} =H_{mean}^f+H_{mean}^b $ with
\begin{eqnarray}
H_{mean}^f  &=&J'\sum_{<ij>} (\psi_i^\dagger U_{ij} \psi_j +c.c.)
+ \sum_{i} \psi_i^\dagger a_0^{(l)}(i) \tau^l \psi_i \nonumber \\
H_{mean}^b  &=&t'\sum_{<ij>}( b_i^\dagger U_{ij} b_j +c.c.)
+ \sum_{i} b_i^\dagger a_0^{(l)}(i) \tau^l b_i
\label{a1}
\end{eqnarray}
where $J'=\frac{3J}{8}$, $t'=\frac{t}{2}$. The fields
$\psi^T=(\psi_{\up} , \psi_{\down}^\dag )$
and
$b^T=(b_1, b_2)$ are $SU(2)$ doublets.
The MF $d$-wave SC state at $T=0$
is described by the following ansatz:
$U^d_{i,i+\hat x} = -\chi \tau^3 - \eta \tau^1$,
$U^d_{i,i+\hat y} = -\chi \tau^3 + \eta \tau^1$
and
$ a_0^{(3)}(i)=a_0$, $a_0^{(1,2)}(i)=0$,
 $\vev{ b^T(i)} =(\sqrt{x} , 0)$.
At higher temperatures ($T>T_c$), boson condensation disappears,
$\langle b(i) \rangle=0$ and $a_0^{(3)}(i)=0$; the above ansatz
describes a normal metallic state with a pseudogap.

The analysis in Ref. \cite{LNNW} indicates that the $SU(2)$ theory
contains a soft mode which correspond to rotation from $b_1$ into $b_2$.
Such a soft mode was overlooked in the $U(1)$
theory. With the hard core repulsion between the bosons, one can show
that the magnitude of quantum fluctuations of bosons is comparable with
the magnitude of the condensation. Thus the  quantum fluctuations have
a potential to completely destroy the boson condensation, or at least
they will reduce the the boson condensation by a finite fraction.
This leads us to consider two scenarios:
\begin{enumerate}
\item \label{scA}
Single boson condensation (SBC) where $\vev{b_1}=\sqrt{x_c}< \sqrt{x}$,
$\vev{b_2} =0$ and a factor $x-x_c$ of bosons remain incoherent and separated
from the condensate by an energy gap $\Del_b$ due to the Higgs mechanism.
\item \label{scB}
Boson pair condensation (BPC) where $\vev{b_\al}=0$ but
$\vev{ b_\al (\v i) b_\bt (\v j)} \neq 0$. We note that BPC is sufficient
to generate electron pair condensation $\vev{c_\up(\v i)c_\down(\v j)}\neq 0$.
The boson pairing will also generate an energy gap $\Del_{bp}$.
\end{enumerate}
We defer a
discussion of the motivation and more detailed formulation of the BPC state,
but remark at the point that we do not know a priori whether SBC or BPC is
favored. Of course neither $\vev{b}$ nor $\vev{bb}$ is gauge invariant and the
only real distinction between these scenarios lies in their experimental
consequences. In particular, we will show that BPC implies $\alpha = 1$ while
SBC implies $\alpha < 1$.

The MF electron propagator is given by
the product of the boson and the fermion propagator:
\begin{equation}
 G_0(\om,{\v k})= \frac i2 \int \frac{d\nu d^2 \v q}{(2\pi)^3}
  {\rm Tr} \v G^{b}(\nu-\omega, \v q-\v k) \v G^f(\nu, \v q)
\end{equation}
where $i\v G^f = \langle \psi \psi^\dag \rangle $ and
$i\v G^{b} = \langle b b^\dag \rangle $ are $2\times 2$ matrices.
In the normal state, the boson propagator for finite $\v A$
satisfies $\v G^b_{\v A}(\om, \v k)= \v G^b(\om, \v k-\frac{e}{c}\v A)$.
Thus the MF electron propagator also shift with $\v A$ as expected:
$G_{0,\v A}(\om,{\v k}) = G_0(\om,{\v k} + \frac{e}{c}\v A)$.
This is a consequence of the gauge symmetry.
However, in the SC state, the bosons may condense according to scenario
\ref{scA} and the
boson propagator contains two terms.
The first term, coming from SBC does not shift with $\v A$.
This is because as we continuously turn on a constant $\v A$, $\vev{ b}$ has
to satisfy periodic boundary condition in a box and cannot be changed (unless
we want to create a vortex). On the other hand,
the second term, coming from non-condensed bosons,
shifts with $\v A$, since the boson dispersion shifts with $\v A$. 
Thus for finite $\v A$, the
boson propagator is given by
 $\v G_{\v A}^{b}(\om, \v k) = -i \vev{ b} \vev{b^\dag} \del(\om, \v k)
 + {\v G^b_{in}}(\om, \v k -\frac{e}{c}\v A)$.
Now it is clear that, in the SC state,
the poles in the MF electron propagator, coming from the product
of the fermion propagator and the first term of the boson propagator,
do not shift with $\v A$.
(In RPA, the generation of the fictitious gauge field $\v a$
by a finite $\v A$ shifts the fermion dispersion by $\frac{e\alpha}{c}\v A$
where $\alpha = x$ as discussed earlier.)

To obtain \Eq{Esc} we have to go beyond the MF theory. First we
assume that not all bosons condense:
$\vev{b_1(\v i)}=\sqrt{x_c}$ with $x_c$ less then
the total boson density $x$. The non-condensed
bosons have small energies and momenta near $\v k=(0,0),\ (\pi,\pi)$,
the two bottoms of the boson band. We ignore the gap
$\Del_b$ for simplicity and model $\Im  {\v G^b_{in}}$ by
peaks of finite width
near $\om=0$ and $\v k=(0,0),\ (\pi,\pi)$.
With those assumption, one can show that the MF electron propagator can
be approximated by \cite{WL,LNNW}
\begin{eqnarray}
 && G_{0,\v A}(\om,\v k)\simeq
 \frac{x_c}{2} \left[\frac{ \left(u^f(\v k)\right)^2 }{\om-E(\v k)-i0^+}
 + \frac{ \left(v^f(\v k)\right)^2 }{\om+E(\v k)-i0^+} \right] \nonumber\\
 &&\ \ +\frac{x-x_c}{2} \left[
   \frac{ \left(u^f(\v k+
   \frac{e}{c}\v A)\right)^2 }{\om-E(\v k+\frac{e}{c}\v A)-i\Ga}
   \right. \nonumber\\
 &&\ \ \ \ \ \left. + \frac{ \left(v^f(\v k+
   \frac{e}{c}\v A)\right)^2 }{\om+E(\v k+\frac{e}{c}\v A)-i\Ga} \right]
 + G_{in} (\om,\v k +\frac{e}{c}\v A)
\end{eqnarray}
where $E=\sqrt{(\eps^f)^2+ (\Del^f)^2}$,
$\eps^f = -2J' \chi(\cos k_x + \cos k_y)+a_0^{(3)}$,
$ \Del^f = -2J' \eta (\cos k_x - \cos k_y)$,
$u^{f} = \frac{1}{\sqrt{2}} \sqrt{1 + \frac{\eps^f}{E}} $,
and $v^{f} = \frac{\Del^f}{\sqrt{2} |\Del^f |}
\sqrt{1 - \frac{\eps^f}{E}} $.
The first term in $G_{0,\v A}$
comes from the SBC and does not shift with
$\v A$. The second term comes from the peak in $\Im {\v G^b_{in}}$ which
shifts
with $\v A$. The decay rate $\Ga$ comes from the finite peak width of
$\Im {\v G^b_{in}}$ in both $\om$ and $\v k$ direction. The last term is the
incoherent part.

When there is BPC,
the electron will have a non-vanishing off-diagonal propagator.
In scenario \ref{scA}, SBC automatically generates BPC
$\vev{b_\al(\v i) b_\bt(\v j)}= x_c \del_{1\al}\del_{1\bt}$.
This yields an off-diagonal electron propagator $-i\vev{ c_\up c_\down }$:
\begin{eqnarray}
  F_{0,\v A} \simeq \frac{x_{pc} u^f v^f}{2}
  \left[\frac{1}{\om-E-i\Ga_b}-\frac{1}{\om+E-i\Ga_b}
  \right]
  \label{mc13}
\end{eqnarray}
where $x_{pc}=x_c$ and $\Ga_b\to 0^+$. More generally, with or without SBC,
the bosons can form pairs with size $l_b$. We shall show later that,
in scenario \ref{scB}, the BPC is characterized by
$\vev{b_\al(\v i) b_\bt(\v j)}= \frac{x_c}{2} f(\v i-\v j)(
\del_{1\al}\del_{1\bt}+(-)^{\v i-\v j}\del_{2\al}\del_{2\bt}) $.
Despite a different form of BPC, the
off-diagonal electron propagator still takes the same form \Eq{mc13} with
$\Ga_b\sim \frac{J}{l_b}$, as long as $l_b$ is much larger than
 the lattice spacing.
Note that $F_{0,\v A}(\om, \v k)$ does not depend on $\v A$, since
the BPC $\vev{b b^T}$ cannot depend on $\v A$.

A second ingredient in going beyond MF theory is to calculate
the electron propagator
through a ladder diagram \cite{WL,LNNW} to include effects of pairing
between the boson and the fermion cause by the $SU(2)$ gauge fluctuations. 
The gauge fluctuations induce the following effective interaction
$\frac13 V \psi^{\dagger} \vec{\tau} \psi b^{\dagger}
\vec{\tau}b
=  V c^{\dagger}c 
 - V \frac{1}{6} \psi^{\dagger} \vec{\tau}b b^{\dagger}\vec{\tau} \psi $ 
with $V>0$.
Here we will only use the first term
$ V(c_\up^\dag c_\up+ c_\down^\dag c_\down)$
because an analytic calculation is possible. The more general interaction
will not modify our results qualitatively.
The resulting electron propagator is given by
\begin{eqnarray}
&& \v G_{\v A}(\om,\v k) \equiv
 \pmatrix{ -i \vev{ c_\up c_\up^\dag} & -i\vev{c_\up c_\down} \cr
  -i \vev{ c_\down^\dag c_\up^\dag} & -i\vev{c_\down^\dag c_\down} \cr}
  \nonumber\\
 &=& \left[
 \pmatrix{
 G_{0,\v A}(\om,\v k) &  F_{0,\v A}( \om, \v k) \cr
 F_{0,\v A}(\om,\v k) & -G_{0,\v A}(-\om,-\v k) \cr
 }^{-1} - V\tau^3 \right]^{-1}
 \label{mc15}
\end{eqnarray}

Let us assume for simplicity $\Ga=\Ga_b=0^+$. We first consider scenario
\ref{scB} where there are no SBC, $x_c=0$. For $\v A=0$, the poles
of $G_{11}(\om \v k)$ comes in pairs of opposite signs, just as in BCS theory.
However the total residue is $\frac{x}{2(1-VG_{in})^2}$, significantly reduced
from the BCS value. There are two positive branches which determine the
qp excitations
\begin{equation}
 E^{(sc)}_\pm (\v k)= \sqrt{ \t E_\pm^2 + \left(\frac{x_{pc}}{x} \Del \right)^2}
 \label{mc16}
\end{equation}
where
\begin{equation}
 \t E_\pm = \pm \sqrt{(\eps-\t \mu)^2+\Del^2
   -\left(\frac{x_{pc}}{x} \Del \right)^2} -\t \mu
 \label{mc17}
\end{equation}
and $\t \mu =-\frac{xV}{4(1-VG_{in})}$.
In order to interpret those results, let us
first consider the normal state which is recovered by setting $x_{pc}=0$ in
\Eq{mc16} and \Eq{mc17}, yielding the normal state dispersion
$E^N_\pm \equiv \t E_\pm(x_{pc}=0)$. This corresponds to
a massless Dirac cone initially centered
at $(\pm \pi/2, \pm \pi/2)$ when $V=0$ which is the MF fermion spectrum
of the staggered-Flux (s-Flux)
phase. The effect of $V$ (the boson-fermion pairing) is
two-fold. The $\t \mu$ inside the square-root shift the location of the node
towards $(0,0)$ by a distance $\Del k=-\t \mu/v_F$ while the last term shift
the spectrum upwards. The cone intersects the Fermi energy to form a small
Fermi pocket with linear dimension of order $x$.
As shown in Fig. 1a, the spectral weight is concentrated on one side of the
cone, so that only a segment of FS on the side close to the origin
carries substantial weight.  This is the origin of the notion of ``FS segment''
introduced in Ref. \cite{WL,LNNW}.
\vskip -.05in
\begin{figure}
\epsfxsize=3.5truein
\centerline{ \epsffile{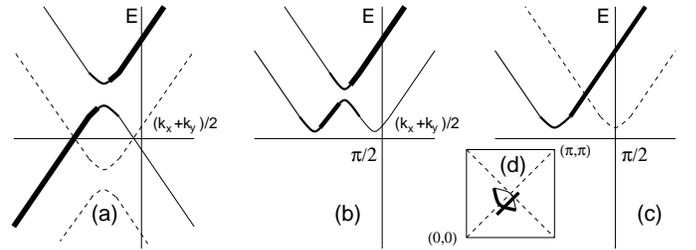} }
\caption{
Schematic illustration of
the qp dispersion (the pole locations of $\v G$) for  (a) normal state,
and SC state with (b) $0<x_{pc}<x$ and (c) $x_{pc}=x$. The line
thickness indicates the size of residue of $G_{11}$,
and dash line indicates vanishing
residue. The momentum scan is along the straight line in (d) where the curved
segment is the FS segment in the normal state.
}
\end{figure}
\vskip -.3in
Now let us see what happens in the SC state when $x_{pc}\neq 0$. \Eq{mc16} takes
the standard BCS form if $\t E_\pm$ is interpreted as the normal state
dispersion. However, $\t E_\pm$ differs from the normal state spectrum
$E_\pm^N$ by the appearance of the term $-(x_{pc}\Del/x)^2$ in \Eq{mc17}.
Close to the node this term is small so that qualitatively the spectrum
develops from the normal state in a BCS fashion, as shown in Fig. 1b. This is
particularly true if the higher energy gap between the two branches is smeared
by lifetime effects. Thus we see that the ``FS segment'' is gapped
in a BCS-like fashion. However, the velocity $v_2$ in the $(1,-1)$ direction,
being proportional to $x_{pc}/x$, does not extrapolate to the gap at
$(0,\pi)$ (which is essentially independent of $x_{pc}$), but cross-over to it
at the edge of the FS segment. It is worth remarking that in the
special case $x_{pc}=x$, $E^{(sc)}_\pm$  reduces to the standard BCS form
with the normal state dispersion $\eps(\v k)$, a chemical potential $2\t \mu$
and a SC gap $\Del(\v k)$.
The high energy gap closes and spectral weight on one branch vanishes,
yielding a BCS spectrum as shown in Fig. 1c.
It is easy to see that \Eq{mc15} is proportional to the BCS Green function in
this special case.

We have also calculated the effect of constant $\v A$ on the qp
dispersion, to linear order of $\v A$. This adds a term
$\frac1c \v j_\pm \cdot \v A$ to \Eq{mc16} where
$\v j_\pm $ is interpreted as the current carried by the qp. We
recall that in standard BCS theory, the current is given in term of the normal
state spectrum by $c \prt_{\v A} \eps_{\v A}=e \prt_{\v k} \eps$ because
$\eps_{\v A} (\v k) = \eps(\v k+ \frac{e}{c} \v A)$.
Remarkably this is almost true
in our case in the sense that $\v j_\pm$ is given by
$c \prt_{\v A} \t E_{\pm, \v A}$, where $\t E_{\pm, \v A}$
is obtained by replacing
$\v k$ by $\v k + \frac{e}{c} \v A$
in $\eps$, $\t \mu$ and $\Del$ everywhere  in
\Eq{mc17} except for the term $\left( \frac{x_{pc}}{x} \Del \right)^2$, which
is kept independent of $\v A$. Near the node, $\Del$ is negligible
so that the current is very close to $e \prt_{\v k} \t E \simeq
e \prt_{\v k} E^N$
(which becomes exactly $e\prt_{\v k} \eps$ along the diagonal),
thus reproducing \Eq{Esc}.
We have checked numerically that
even away from the node in the region of the ``FS segment'',
the current is remarkably close to $e \prt_{\v k} E^N$,
which can be quite different from the BCS value
$e \prt_{\v k} \eps$ near the edge of the FS segment.

Next we briefly comment on what happens under scenario \ref{scA}. The main
difference is that the current carried by qp is no longer equal to
$e \v v_F$, but reduce from it depending on $x_c$. It is clear that in the
extreme case of $x-x_c=G_{in}=0$, $G_{0 \v A}$ and $F_{0\v A}$, and hence
$\v G_{\v A}$ do not depend on $\v A$, so that $\v j=0$. As $x_c$ varies from
$x$ to 0,  $\alpha$ interpolates between $0$ and
1. (Note $\v v_F$ is defined as the normal state Fermi velocity
in the $(1,1)$ direction, $\v v_F= \prt_{\v k} E^N$.
It is also exactly equal
to the qp velocity in the $(1,1)$ direction at the SC nodes.)
Due to strong
quantum fluctuations of bosons, $x-x_c$ is of order $x$ and hence
$\alpha$ is order unity. The main question is whether $\alpha$ is exactly $1$.
According to our model, whether $\alpha=1$ depends on whether there is a
SBC. Thus it is very interesting and important
to determine $\alpha$ experimentally.

From \Eq{Eq.3}, the temperature dependence of the London
penetration depth gives a direct measurement of $\alpha^2 \frac{v_F}{v_2}$.
Density of
states measurements using the $T^2$ coefficient of the specific heat yields
$v_F v_2$. The Fermi velocity can be estimated from transport
measurements\cite{Ong} or
high resolution photoemission experiment. Thus in principle the quantities
$\alpha$, $v_F$ and $v_2$ can be measured. It is of course of great interest to
establish how close $\alpha$ is to $1$, or whether $v_2$ is reduced with respect
to that extrapolated from the energy gap at $(0,\pi)$ measured by
photoemission or tunneling. 

We may regard $\alpha=1$ as a signature of spin-charge recombination,
{\it ie} the boson and fermion binds (through the ladder diagram) into an
electron which responds fully to $\v A$. We have so far focussed our
discussion to low energy excitations near the nodes. At higher energy away
from the Fermi surface, the binding may become unimportant and the electron
spectrum is given by the convolution of the fermion and boson spectrum. In the
BPC state, an energy gap $\Delta_{bp}$ arises in the boson spectrum, which
should lead to a shift of the electron spectral function in the SC state
relative to the normal state by the energy $\Delta_{bp}$ towards higher
binding energy.

Finally we comment on finite temperature behaviors. In addition to the
reduction of superfluid density due to thermal excitation of qp,
\cite{LeeWen} we expect $x_{pc}$ to decrease with increasing $T$, leading to a
reduction of $v_2$: $v_2(T)=\frac{x_{pc}(T)}{x_{pc}(0)}v_2(0)$.
As $T$ reaches $T_c$, $x_{pc}=v_2=0$ and the nodes of $E^{(sc)}$ become the
``FS segment'' while the spin gap near $(0,\pi)$ remain finite. We
see that $x_{pc}$ plays the role of the order parameter of the transition,
so that
the temperature dependence of $x_{pc}$
is described by a Ginzburg-Landau theory 
near the transition.

We complete our discussion by giving a more microscopic motivation
for the notion of BPC. It was pointed out recently \cite{NL} that one way of
capturing the physics of strong boson fluctuation is to attach flux tube of
opposite sign to $b_1$ and $b_2$ (in s-Flux gauge), converting them to
fermions. This has the advantage that at the MF level, time reversal
symmetry is not broken. In the s-Flux gauge, the massless gauge field is
simply a $U(1)$ gauge field which couple to $b_1$ and $b_2$ with opposite
gauge charge. This problem was treated by Bonesteel {\it et al}
\cite{B} who found that
there is a instability towards boson pairing with $\vev{b_1 b_2}\neq 0$.
Thus it is
natural to assume that, in the s-Flux gauge, only $\vev{b_1(\v i)b_2(\v j)}$
is non-zero in the BPC. After transformed to the $d$-wave
gauge, $\vev{b b^T}$ becomes the one that we used below \Eq{mc13}.
We expect the energy gap $\Delta_{bp}$ to scale with the effective ``Fermi''
energy, {\it ie} $x$.

In summary we develop a theory for quasiparticles in the high $T_c$
superconducting state. Our theory reproduces both small superfluid density
$\rho_s \sim x$ and large $\alpha \sim 1$, as required by experiments.

PAL acknowledges support by NSF-MRSEC Grant
No. DMR--94--00334.  XGW is supported by NSF Grant No. DMR--97--14198.

\end{multicols}

\end{document}